\documentclass[12pt]{iopart}

\usepackage{cite}
\usepackage[normalem]{ulem} 
\usepackage{soul} 
\usepackage{graphicx}
\usepackage{dcolumn}
\usepackage{bm}
\usepackage{euscript,stmaryrd}
\usepackage{xcolor}
\usepackage[utf8]{inputenc}
\usepackage[T1]{fontenc}
\usepackage{mathptmx}
\usepackage{epsf,amsmath,amssymb,verbatim,color,multirow,pifont,mathrsfs,soul,amsthm, wasysym,bm,amsbsy}
\usepackage{graphicx, upgreek}
\usepackage[us,12hr]{datetime}
\usepackage{xcolor}
\usepackage{arydshln}
\usepackage{hyperref}
\usepackage{arydshln}
\usepackage{graphicx}
\usepackage{mathtools}
\usepackage[symbol]{footmisc}

\theoremstyle{definition}

\newcommand\redout{\bgroup\markoverwith
{\textcolor{red}{\rule[0.5ex]{2pt}{0.8pt}}}\ULon}

\newmuskip\pFqmuskip

\newcommand*\pFq[6][8]{%
  \begingroup 
  \pFqmuskip=#1mu\relax
  \mathcode`\,=\string"8000
  \begingroup\lccode`\~=`\,
  \lowercase{\endgroup\let~}\pFqcomma
  {}_{#2}F_{#3}{\left[\genfrac..{0pt}{}{#4}{#5};#6\right]}%
  \endgroup
}
\newcommand{\pFqcomma}{\mskip\pFqmuskip}

\usepackage{fancyhdr}
\usepackage{hyperref}
\hypersetup{
    colorlinks,
    citecolor=blue,
    filecolor=blue,
    linkcolor=blue,
    urlcolor=black
}

\usepackage{blindtext}
\usepackage{titlesec}

\begin{document}

\title{Hopf-Induced Desynchronization}

\author[1]{Seungjae Lee$^*$}
\address{Chair for Network Dynamics, Institute of Theoretical Physics and
  Center for Advancing Electronics Dresden (cfaed), Technische Universit\"at Dresden, 
  01062 Dresden, Germany}

\ead{seungjae.lee@tu-dresden.de}

\author[2]{Lennart J. Kuklinski}
\address{Chair for Network Dynamics, Institute of Theoretical Physics and
  Center for Advancing Electronics Dresden (cfaed), Technische Universit\"at Dresden, 
  01062 Dresden, Germany}

\author[3]{Moritz Th\"umler}
\address{Chair for Network Dynamics, Institute of Theoretical Physics and
  Center for Advancing Electronics Dresden (cfaed), Technische Universit\"at Dresden, 
  01062 Dresden, Germany}

\author[4]{Marc Timme$^*$}
\address{Chair for Network Dynamics, Institute of Theoretical Physics, 
  Center for Advancing Electronics Dresden (cfaed), and Center Synergy of Systems, Technische Universit\"at Dresden, 
  01062 Dresden, Germany}
\address{
  Lakeside Labs, Lakeside B04b, 9020 Klagenfurt, Austria
}

\ead{marc.timme@tu-dresden.de}

\vspace{10pt}
\begin{indented}
\item[]\today
\end{indented}

\begin{abstract}
The emergence of synchrony essentially underlies the functionality of many systems across physics, biology and engineering. In all established  synchronization phase transitions so far, a stable synchronous state is  connected to a stable incoherent state: For continuous transitions, stable synchrony directly connects to stable incoherence at a critical point, whereas for discontinuous transitions, stable synchrony is connected to stable incoherence via an additional unstable branch.
Here we present a novel type of transition between synchrony and incoherence where the synchronous state does not connect to the state of incoherence. We uncover such transitions in the complexified Kuramoto model with their variables and coupling strength parameter analytically continued. Deriving a self-consistency equation for a quaternion order parameter that we propose helps to mathematically pin down the mechanisms underlying this transition type. 
Local numerical analysis suggests that the transition is linked to a Hopf bifurcation destabilizing synchrony, in contrast to branching point bifurcations established for the transition between synchrony and incoherence so far.
\end{abstract}

\newpage
\section{\label{sec:Intro}Introduction}

Self-organized synchronization, the temporal coordination of states of coupled units, represents a fundamental ordering process emerging in nonlinear dynamical systems~\cite{pikovsky2001universal,strogatz2012sync,acebron2005kuramoto,Rodrigues2016kuramoto}. Examples range from phenomena in nature such as the collective flashing of fireflies, the coordinated activity of pacemaker cells in the heart or spiking neurons in the brain~\cite{Ermentrout1991,wang2013,Ermentrout_Kopell1991,KISS20181,Bick2020_review} to those underlying the function of engineered systems such as phase-locking in AC electric power grids, the self-organization of pedestrian stepping dynamics or of drone communication ~\cite{Strogatz2005_bridge,Schilcher2020,Schilcher2021,witthaut2022collective2}.
With increasing coupling among the units, asynchronous, incoherent collective states are replaced by system-wide synchrony with strongly aligned states. In the thermodynamic limit, a phase transition emerges with asynchronous states below a critical coupling strength and (partially) synchronous states above it. 
Synchronization phase transitions have been originally found to be continuous such that the degree of temporal order (synchrony) gradually increases starting from complete disorder (asynchrony)~\cite{winfree1967biological,Kuramoto1974nonLinearOscillator,Strogatz1991}. 

Half a century ago, Kuramoto \cite{Kuramoto1974nonLinearOscillator} introduced a paradigmatic model for the emergence of synchrony among limit-cycle oscillators. The simplicity and partial analytic accessibility of the Kuramoto model made it the Drosophila of studies on synchronization and temporal coordination processes~\cite{acebron2005kuramoto,Rodrigues2016kuramoto}. In particular, it has opened up a huge field for exploring synchronization phase transitions of coupled oscillatory units.  Specifically, Kuramoto studied a system of $2\pi$-periodic phase variables $x_\mu \in \mathbb{S}^1 = \mathbb{R}/2\pi\mathbb{Z}$ for $\mu \in [N]:=\{1,2, \cdots, N\}$ and found that the absolute value of the order parameter~\cite{Strogatz2000kuramotoCrawford,kuramoto2003chemical}
\begin{flalign}
    r e^{\textrm{i}\Psi} = \frac{1}{N}\sum_{\nu=1}^N e^{\textrm{i}x_\nu} = \frac{1}{N}\sum_{\nu=1}^N (\textrm{i}\sin x_\nu + \cos x_\nu) \in \mathbb{C} \label{eq:OP_kuramoto}
\end{flalign} 
that quantifies the degree of synchrony of the $x_\mu$, gradually starts increasing from zero once the coupling strength increases beyond a critical value. 
Analytic work for oscillators with frequencies drawn from a unimodal distribution confirmed the continuous nature of the phase transition to synchrony. Later work has demonstrated continuous synchronization transitions across many oscillatory systems. Some work has shown that discontinuous phase transitions may equally occur in different instances~\cite{PhysRevLett.109.164101,Skardal2020,millan2020},  for bimodal or compact-support frequency distributions~\cite{bimodal,pazo_discon} and in certain networks with power-law degree distributions~\cite{PhysRevLetters106.128701,BOCCALETTI20161,d2019explosive,liu2024early} as well as in experimental studies, e.g., Belousov-Zabotinsky (BZ) oscillatory reactions~\cite{Dumitru2020}. 

So far, established discontinuous synchronization transitions have been viewed to arise from a bistability between incoherence and synchrony. They occur due to a shift in the critical nature of a branching point bifurcation
(such as a pitchfork or a transcritical bifurcation) from supercritical to subcritical, as theoretically consolidated recently by Kuehn and Bick~\cite{kuehn2021}. More precisely, for a continuous synchronization transition, stable synchrony directly bifurcates off the incoherent state at a \textit{supercritical} branching point. Upon varying a secondary parameter, a discontinuous transition emerges where a state of unstable synchrony bifurcates off the incoherent state at a \textit{subcritical} branching point and then becomes stable via a saddle-node bifurcation (also known as fold bifurcation, see Sec.~\ref{sec:discussion}), implying that an unstable branch connects the stable incoherent state and stable synchronous states. Therefore in both, the established continuous and discontinuous transitions between synchrony and incoherence, the incoherent and the synchronous states are connected to each other, either directly or via an additional unstable branch.

In this article, we uncover a novel type of phase transition to synchrony: A discontinuous phase transition where incoherent and synchronized states are detached from each other in state space and parameter space such that there exists no connecting branch bridging them. 
We mathematically identify these discontinuous transitions in analytically continued Kuramoto models exhibiting complex instead of real state variables and coupling parameters, compare \cite{Moritz_thuemler2023, lee_complex,Lee2025}. We introduce a \textit{quaternion order parameter} $q\in\mathbb{H}$ that generalizes the standard (complex) Kuramoto order parameter $r e^{\textrm{i}\Psi}$ defined in Eq.~\eqref{eq:OP_kuramoto}. 
The analytical insights resulting from the self-consistency equations for $q$ explicate two intriguing findings. First, the synchronous state is indeed disconnected from the incoherent state, in contrast to known discontinuous transitions, where unstable branches create a bridge between them. Second, the phase transition is almost always discontinuous, such that discontinuous transitions prevail across parameter space, supporting numerical observations,
in stark contrast to the 
balance between continuous and discontinuous transitions established so far \cite{kuehn2021}. Finally, local numerical analysis suggests that a Hopf bifurcation, rather than the commonly established saddle-node bifurcation, destabilizes the collective state of synchrony.

\begin{figure}[t!]
\centering
\includegraphics[width=0.7\columnwidth]{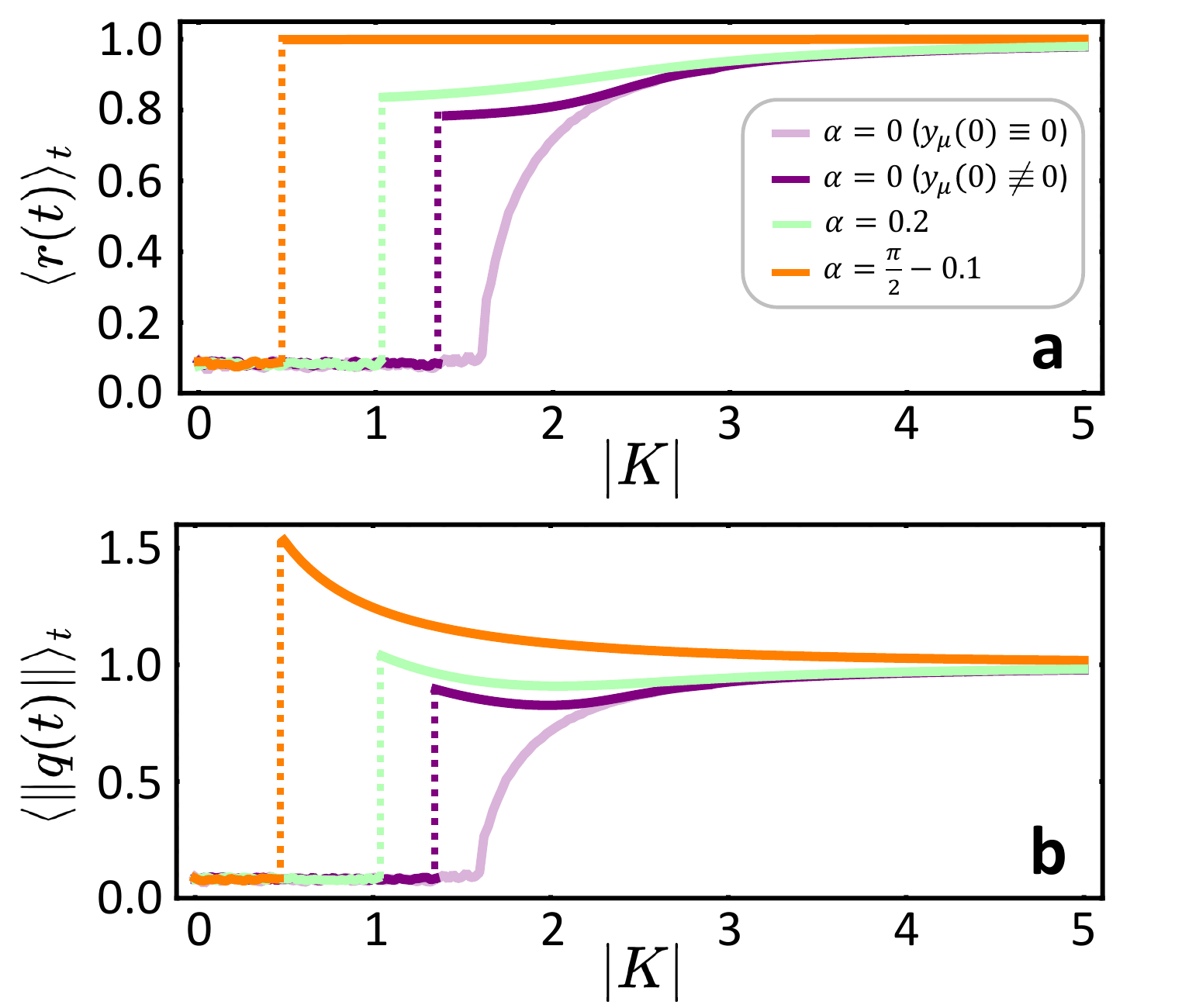}
\caption{\label{Fig:fig1} \textbf{Quaternion order parameter characterizes transitions to synchrony.} (a) The classical Kuramoto order parameter \eqref{eq:OP_kuramoto} as a function of the coupling strength $|K|$ for different $\alpha$. (b) The quaternion order parameter \eqref{eq:QOP} as a function of the coupling strength $|K|$ for the same $\alpha$ as in (a). Light purple curves: $\alpha = 0$ with $y_\mu(0) \equiv 0$ for all $\mu \in [N]$, showing a continuous transition to synchrony in the classical Kuramoto model; dark purple curve: $\alpha = 0$ with general initial conditions $y_\mu(0) \not \equiv 0$; light green curve: $\alpha = 0.2$; orange curve: $\alpha = \frac{\pi}{2} - 0.1$.).
 $N=128$ and natural frequencies  randomly symmetrically drawn from Gaussian density for all $\alpha$.}
\end{figure}

\section{\label{sec:}Quaternion order parameter distinguishes synchrony from incoherence}

Let us consider the dynamics of networks of units with complex variables $z_\mu = x_\mu +\textrm{i} y_\mu$, with $x_\mu \in \mathbb{S}^1$  and  $y_\mu  \in \mathbb{R}$, coupled through sinusoidal interaction functions, thereby analytically continuing the original Kuramoto model ~\cite{Moritz_thuemler2023,lee_complex,Lee2025}.
The coupling constant is also continued to be complex, $K = |K|e^{\textrm{i}\alpha} \in \mathbb{C}$   with $\alpha  \in (-\frac{\pi}{2},\frac{\pi}{2})=:\mathbb{P}$.
The network dynamics satisfies
\begin{flalign}
    \frac{\textrm{d}}{\textrm{d} t} z_\mu = \omega_\mu + \frac{K}{N}\sum_{\nu=1}^{N}\sin(z_\nu-z_\mu) \label{eq:CKM}
\end{flalign} 
for $\mu \in [N]$. To ease derivations of our main results, we symmetrically draw pairs of natural frequencies $\omega_\mu=-\omega_{\mu + \frac{N}{2}} \in \mathbb{R}$ for $\mu \in [N/2]$ from (i) an even probability density with $g(\omega)=g(-\omega)$, that induces (ii) a zero mean, $\langle \omega \rangle:=\int_\mathbb{R} \omega g(\omega)\textrm{d}\omega =0$, is (iii) peaked at zero with $g(0) \neq 0$, $g'(0)=0$, and $g''(0) <0$ and exhibits (iv) infinite, monotonically decaying tails, $g(\omega) \rightarrow 0$ together with $\omega g'(\omega)<0$ as $\omega \rightarrow \pm \infty$.

Direct numerical simulations starting from random initial conditions~\cite{initial_condition} intriguingly revealed discontinuous transitions to synchrony for \textit{all} $\alpha \in \mathbb{P}$, see Fig. \ref{Fig:fig1}a. The traditionally found continuous transition emerges only if $\alpha=0$ exactly and in addition the initial values are all chosen in the real invariant subset, i.e., $z_\mu(0) \in \mathbb{R}$ for all $\mu \in [N]$.

Why are discontinuous transitions so persistent?
How does such a transition differ from established explosive synchronization? To address these questions, we first 
distinguish the degree of coherence from incoherence 
by introducing a \textit{quaternion order parameter}
\begin{flalign}
    q &:= \frac{1}{N}\sum_{\nu=1}^N (\textrm{j} ~\overline{\sin z_\nu}+ \cos z_\nu) \notag \\
    &= q_0 + q_1 \textrm{i} + q_2 \textrm{j} + q_3 \textrm{k}\in \mathbb{H}, \label{eq:QOP}
\end{flalign} where $\textrm{i}, \textrm{j}, \textrm{k}$ are basis elements of quaternions $\mathbb{H} $ and an overline indicates complex conjugation~\cite{voight2021quaternion}. Here, $q_0 =: \textrm{Re}(q) \in \mathbb{R}$ is the real part of $q$ and the $q_n \in \mathbb{R}$ for $n\geq 1$ are the vector part components (\ref{appendA}). In the same way as the original (complex) order parameter $r$ in \eqref{eq:OP_kuramoto} effectively decouples the equations of motion of the original Kuramoto model, the quaternion order parameter~\eqref{eq:QOP} decouples the analytically continued equations of motion \eqref{eq:CKM} such that (\ref{appendb})
\begin{flalign}
    \frac{\textrm{d}}{\textrm{d}t}z_\mu = \omega_\mu -K (q_0 +  \textrm{i} q_1) \sin z_\mu + K (q_2 + \textrm{i} q_3) \cos z_\mu  \label{eq:decoupled_CKM}
\end{flalign} for $\mu \in [N]$.
More specifically, Eq.~\eqref{eq:decoupled_CKM} elucidates that each unit variable $z_\mu$ is driven only by itself and the mean-field $q$ of all other units. Beyond this decoupling, the norm $\|q\|:=\left(\sum_{n=0}^3 q_n^2\right)^{1/2}$ measures the degree of synchrony among the units and directly maps to the classical order parameter $r$, eqn.~\eqref{eq:OP_kuramoto} (\ref{appenC}). Uncoupled units yield an incoherent state, with $\|q\|=0$ (see below) while $\|q\|=1$ marks complete synchrony with $(x_\mu, y_\mu) = (x_\nu, y_\nu)$ for all $\mu,\nu \in [N]$. We remark that the unboundedness of imaginary variables may result in $\|q\| > 1$.
Despite this difference, the quaternion order parameter well distinguishes collective dynamical states, especially synchrony and asynchrony, and  characterizes different types of transitions to synchrony, compare Fig.~\ref{Fig:fig1}b to Fig.~\ref{Fig:fig1}a (\ref{appenC}).

\section{\label{sec:self_consist}Self-consistency equation for locked states}

Fixed point solutions of \eqref{eq:CKM} determine generalized locked states with temporally fixed $z^*_\mu$, both for finite-size systems and in the thermodynamic limit. For finite $N$, such a complex locked state is represented by an algebraic equation 
\begin{equation}
    0 = \omega_\mu - K q^* \sin z^*_\mu
\end{equation}
implying $z^*_\mu = \sin^{-1} \frac{\omega_\mu}{K q^*}$  for  $\mu \in [N]$.
As the complex sine function is entire, such a solution exists for all parameters $K\in\mathbb{C}\setminus\{0\}$, all oscillators are locked and there is no drifting oscillator. Here, we exploited that the symmetric natural frequencies restrict $q^*$ to a complex subspace of $\mathbb{H}$ such that $q^*_2 = q^*_3 = 0$ in \eqref{eq:QOP}. 
In the thermodynamic limit $N\rightarrow\infty$, such locked states become
\begin{flalign}
    z^*(\omega) =  \sin^{-1} \bigg( \frac{\omega}{K q^*} \bigg) \in \mathbb{C}, \label{eq:complex_locked_state_thermo}
\end{flalign} yielding the self-consistency condition
\begin{flalign}
    q^* &= \int_\mathbb{R} \bigg(\textrm{j}\overline{\sin z^*(\omega)} + \cos z^*(\omega) \bigg) g(\omega) \textrm{d}\omega \notag \\
    &=\int_\mathbb{R}  \bigg(1-\frac{\omega^2}{K^2q^{*2}}\bigg)^{\frac{1}{2}} g(\omega) \textrm{d} \omega . \label{eq:self_consis_QOP}
\end{flalign} 
For simplicity of presentation, we hereafter omit the asterisk symbol.

\section{\label{sec:continu_sync}Continuous synchronization transition} 

The original Kuramoto model ($\alpha=0$) evolves on an invariant manifold 
$\mathbb{R}^N\times\{0\}^N$ 
if both the initial conditions $z_\mu(0)\in\mathbb{R}$ and the coupling constant $K \in \mathbb{R}$ are exactly real. For $\alpha = 0$ and $z_\mu(0)\in\mathbb{R}$ for all $\mu \in [N]$,
the quaternion order parameter becomes real, $q \in \mathbb{R} \times \{0\}^3 \subset \mathbb{H}$ due to the symmetric natural frequencies. To yield a self-consistent real $q$, the argument of the root in the self-consistency equation \eqref{eq:self_consis_QOP} needs to be non-negative, thereby constraining the range of frequencies $\omega$ such that
\begin{flalign}
    q &= \int_{-Kq}^{Kq} \sqrt{1-\frac{\omega^2}{K^2 q^2}} g(\omega) \textrm{d}\omega \notag \\
    &= Kq \int_{-1}^{1} \sqrt{1-x^2} g(K q x) \textrm{d}x  .\label{eq:self_con_real}
\end{flalign} 
For locked states ($q>0$), 
we divide \eqref{eq:self_con_real} by $q$ and obtain the critical coupling strength at which a positive order parameter ($q>0$) continuously bifurcates off from the $q=0$ solution representing the incoherent state. As $q \rightarrow 0^+$, we obtain $1 = K_c g(0) \int_{-1}^{1}\sqrt{1-x^2}\textrm{d}x = K_c g(0) \frac{\pi}{2}$, determining $K_c = \frac{2}{\pi g(0)}$. We thereby recover the continuous phase transition to synchrony known from the original Kuramoto model~\cite{Strogatz2000kuramotoCrawford}.

\section{\label{sec:discontinu_sync}Synchrony detached from incoherence}

Intriguingly, we observe a type of discontinuous phase transition that is not only fundamentally different from established continuous but also from established discontinuous transitions. In the following, we present three sets of evidence supporting that the synchronous state is neither directly nor indirectly connected to the incoherent state and that the phase transition is persistently discontinuous across parameter space.

\subsection{Quaternion order parameter}
\begin{figure}[t!]
\centering
\includegraphics[width=0.8\columnwidth]{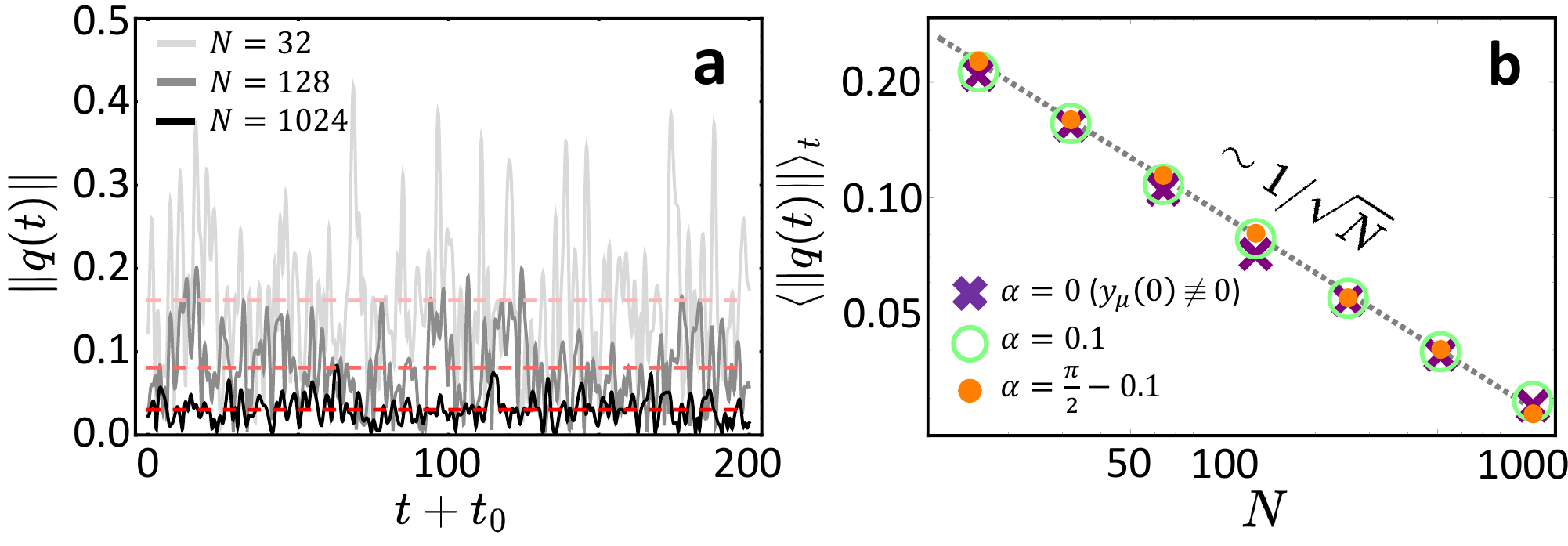}
\caption{\label{Fig:fig2} \textbf{In the incoherent state, the quaternion order parameter decays to zero as $N\rightarrow\infty$.} (a) The quaternion order parameter as a function of time for different system sizes $N=32, 128$ and $1024$, with $\alpha = 0.1$ and $t_0 = 1000$. The red dashed line indicates time averages $\langle \|q(t)\| \rangle_t$ for each system size. (b) The time-averaged quaternion order parameter as a function of system size $N$, shown in log-log scale for different values of $\alpha$ (same color code in Fig.~\ref{Fig:fig1} for different $\alpha$). The dashed guideline indicates $\mathcal{O}(1/\sqrt{N})$ as $N \rightarrow \infty$. Here, $|K|=0.2$ for both panels. }
\end{figure}

To see this, let us consider general $\alpha \neq 0$ or $\alpha=0$ and generic initial conditions $z_\mu \in \mathbb{C}$ such that $q\in\mathbb{C}\setminus\mathbb{R}$ in \eqref{eq:self_consis_QOP}. In the incoherent state, the probability density in the thermodynamic limit is inversely proportional to the oscillator velocity $\frac{\textrm{d}}{\textrm{d}t}z_\mu$ given by \eqref{eq:decoupled_CKM} such that (\ref{appenB234})
\begin{flalign}
    &\rho_\textrm{inc}(x,y,\omega) = \rho_\textrm{inc}(x+\pi,y,-\omega) 
    \label{eq:density_symmetry} \\
    &=  \frac{C(\omega)}{
     \big|\omega -K [(q_0 +  \textrm{i} q_1) \sin(x+\textrm{i}y) +  (q_2 + \textrm{i} q_3) \cos(x+\textrm{i}y)] \big|  } 
     \label{eq:density_incoherent} 
\end{flalign} 
where $C(\omega)$ is a normalization constant. In analogy to the real model~\cite{Strogatz2000kuramotoCrawford}, the symmetry \eqref{eq:density_symmetry} results in  $q_\textrm{inc}=0$ in the incoherent state (\ref{appenB234}). 
Numerical experiments indicate that the quaternion order parameter vanishes for incoherently drifting oscillators in the thermodynamic limit $N \rightarrow \infty$ (Fig.~\ref{Fig:fig2}).

In contrast, the synchronous state exhibits solutions $q$ to \eqref{eq:self_consis_QOP} with  $\|q\|>0$ which is completely \textit{disconnected} from the $\|q\|=0$ solution for the incoherent state as we demonstrate next. Assume they were connected and continuously branched off the $\|q\|=0$ solution. For small $\|q\| \ll 1$, given the properties (i-iv) above,  we write  the natural frequency density as $g(\omega) = e^{f(\omega)}$ and observe that as $\|q \| \rightarrow 0^+$, only small values of $\omega$ contribute.
We asymptotically \cite{timme2020disentangling} obtain $g(\omega) = e^{f(\omega)} \sim g(0) e^{\frac{ f''(0)}{2} \omega^2} $ in the integrand as $\|q \| \rightarrow 0^+$, up to transcendentally small  terms (t.s.t.), so that we can exploit an asymptotic approximation of Laplace-type~\cite{holmes2012introduction,bender1999advanced}. 
For instance, for a Cauchy-Lorentz distribution, $g(\omega) = \frac{1}{\pi} \frac{\Delta}{\omega^2 + \Delta^2} \sim e^{\log(\frac{1}{\pi \Delta}) - \frac{\omega^2}{\Delta^2} +\mathcal{O}(\omega^4)}$ as $|\omega| \rightarrow 0$, while for a Gaussian (normal) distribution we exactly have $g(0) = \frac{1}{\sqrt{2\pi}}$ and $f''(0)=-1$.

Substituting the general asymptotic expansion into \eqref{eq:self_consis_QOP}, we obtain for $Kq \in \mathbb{C} \setminus \mathbb{R}$ 
\begin{flalign}
     q &= g(0) \int_\mathbb{R} \bigg(1-\frac{\omega^2}{K^2 q^2}\bigg)^{\frac{1}{2}} e^{\frac{f''(0)}{2} \omega^2} \textrm{d} \omega + \textrm{t.s.t.} \notag \\
     &= g(0) \frac{\textrm{2}}{|f''(0)|}\bigg( \frac{-1}{K^2 q^2} \bigg)^{\frac{1}{2}} + \mathcal{O}(\|q\|) + \textrm{t.s.t.} \label{eq:contradiction}
\end{flalign} as $\|q\| \rightarrow 0$. The result \eqref{eq:contradiction} implies a contradiction: While the LHS converges to zero asymptotically as $\|q\| \rightarrow 0$, the leading term of the RHS diverges. Hence, the  self-consistency equation does not admit a solution with arbitrary small $\|q\|$. The finding indicates that synchronous states do not connect to the incoherent state.
Any phase transition to synchrony is thus necessarily discontinuous without any branch that bridges synchrony and incoherence (see \ref{appenBbbbb} for more details).

For instance, for a Gaussian distribution $g(\omega)= \frac{1}{\sqrt{2\pi}} e^{-\frac{1}{2} \omega^2}$, the self-consistency equation \eqref{eq:self_consis_QOP} reads
\begin{flalign}
    q &= \frac{2\textrm{i}}{\sqrt{2\pi}}  \bigg( \frac{\pi}{K^2 q^2} \bigg)^{\frac{1}{2}}  U\left(-\frac{1}{2}, 0, -\frac{1}{2} K^2 q^2\right)  \label{eq:gaussian_QOP_self_consist}
\end{flalign} 
with confluent hypergeometric function $U$~\cite{arfken2013mathematical} provided that $Kq \in \mathbb{C} \setminus \mathbb{R}$. The relation \eqref{eq:gaussian_QOP_self_consist} implicitly determines the quaternion order parameter for complexified synchrony (whether it is stable or unstable) as well as illustrates some of its properties, see again Fig.~\ref{Fig:fig1}b and Fig.~\ref{Fig:fig3}. First, $\|q\|$ does not approach zero in any locked state for any $|K|>0$; thus, discontinuous transitions emerge persistently for all $\alpha\in\mathbb{P}$ and without a bridging connection between synchrony and incoherence. Second, $q \rightarrow  1 $ as $|K| \rightarrow \infty$, and thus the quaternion order parameter $\|q\|$ converges to the classical Kuramoto order parameter $r$ as the coupling strength $|K|$ increases.

\subsection{Kuramoto order parameter}

\begin{figure}[ht!]
\centering
\includegraphics[width=0.6\columnwidth]{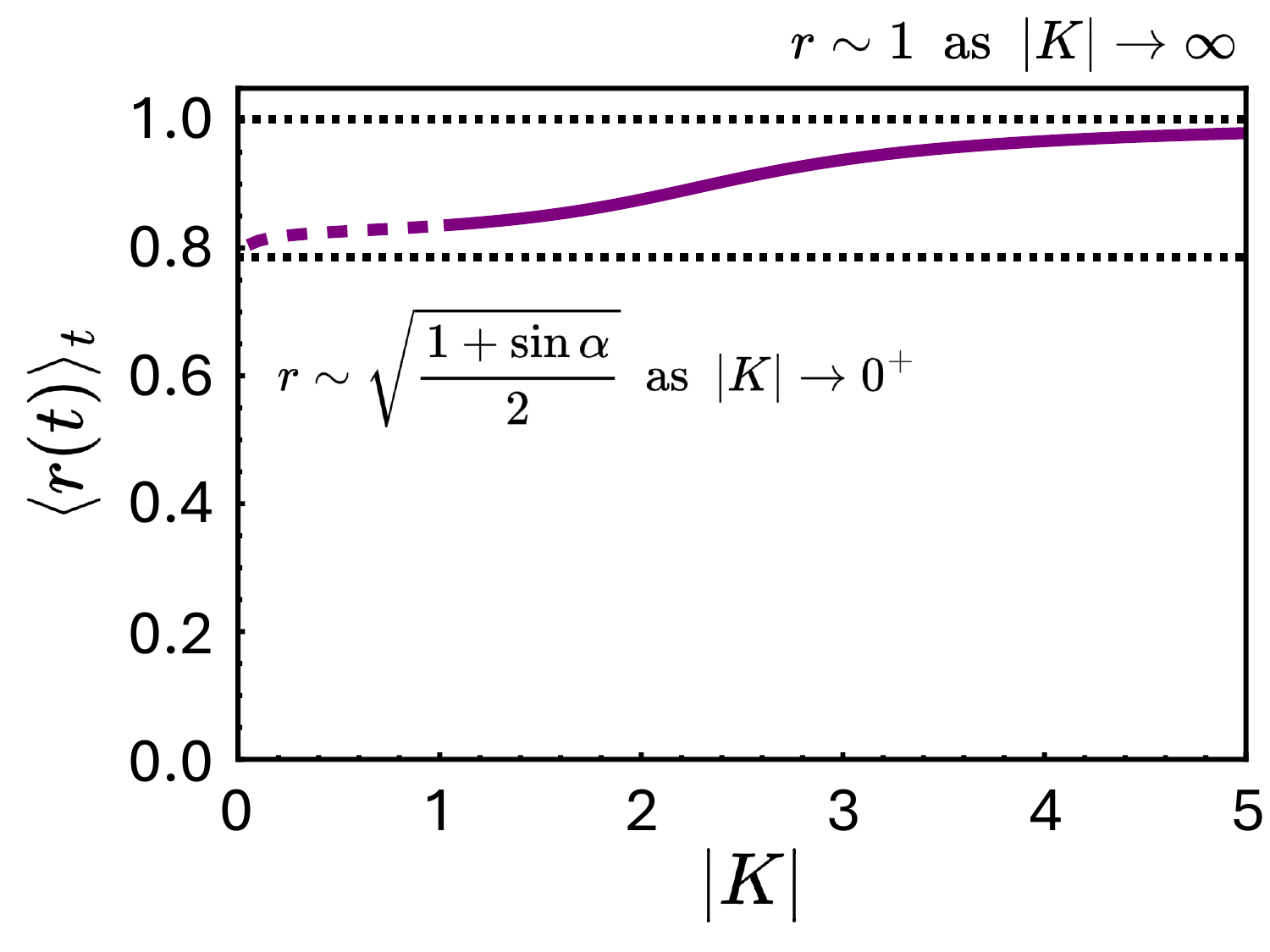}
\caption{\label{Fig:Order_parameter} \textbf{The Kuramoto order parameter is disconnected to the incoherence.} The Kuramoto order parameter is shown as the coupling strength $|K|$ varies. Here, we use $N=128$ and $\alpha=0.2 >0$. The dashed part of the purple curve indicates unstable synchrony for small $|K|$ while the solid part of the curve shows stable synchrony. The below-dotted line indicates the order parameter limit as $|K| \rightarrow 0^+$ whereas the above-dotted line shows the order parameter as $|K| \rightarrow \infty.$
}
\end{figure}

The Kuramoto order parameter also  suggests \textit{disconnectedness} between the synchronized state and the incoherent state (Fig.~\ref{Fig:Order_parameter}). An asymptotic analysis on the Kuramoto order parameter (\ref{appenG}) confirms 
\begin{flalign}
    r \sim \left\{ 
    \begin{array}{cl}
        \sqrt{\frac{1+\sin\alpha}{2}}=\mathcal{O}(1) & \textrm{for} ~~ |K| \rightarrow 0^+ \\
        \\
        1 & \textrm{as} ~~ |K| \rightarrow \infty  
    \end{array} \right. \label{eq:asym} 
\end{flalign} For a small $|K|$, the Kuramoto order parameter for unstable synchrony already assumes values of $\mathcal{O}(1)$ rather than $\mathcal{O}(\frac{1}{\sqrt{N}})$ as $N \rightarrow \infty$ for the incoherent state. As $|K| \rightarrow \infty$, the Kuramoto order parameter asymptotically approaches unity that characterizes a completely synchronized state. Thus, unless the Kuramoto order parameter $r$  varies strongly non-monotonically as a function of $|K|$ (for which we have no numerical support), $r=\mathcal{O}(1)$ does not touch the value $r=0$ of the incoherent state for any $|K|>0$. \footnote{We do not consider the single potential exception at $\alpha=3\pi/2\not\in\mathbb{P}$ that reflects repulsive coupling.}

\subsection{Instability via Hopf bifurcation}

\begin{figure}[ht!]
\centering
\includegraphics[width=0.8\columnwidth]{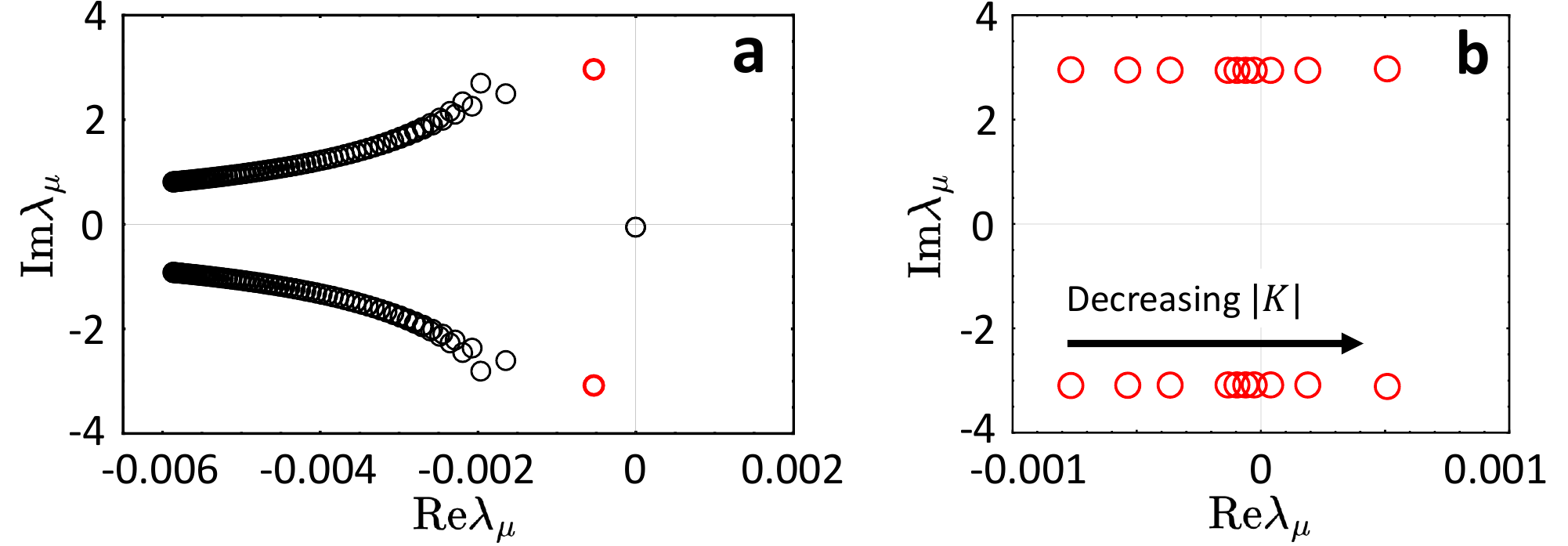}
\caption{\label{Fig:eigenvalue} \textbf{Synchrony destabilizes via a Hopf bifurcation.} (a) Eigenvalues (circles) of a Jacobian matrix in a complex plane indicate stable complexified synchrony at $|K| = 0.6$ (here,  $N=256$, $\alpha=\frac{\pi}{2}-0.1$). The one pair of complex conjugate eigenvalues with largest real part (excluding $\lambda=0$ that exists due to phase shift invariance) is indicated by red circles. (b) As the coupling strength $|K|$ decreases, a pair of complex conjugate eigenvalues crosses over the imaginary axis to positive real parts, indicating destabilization of synchrony.
}
\end{figure}

Finally, a numerical stability analysis of the complex locked state suggests that complex locked states loose their stability in a Hopf bifurcation.

The existence of synchrony in the form of a complex locked state $\mathbf{z}^*$ for all parameters $K\neq 0$ enables us to study its local stability by numerically evaluating the eigenvalues of the local Jacobian at $\mathbf{z}^*$, see Figure~\ref{Fig:eigenvalue}.  Figure~\ref{Fig:eigenvalue}a at $|K|=0.6$ indicates stable synchrony. As $|K|$ decreases (Figure~\ref{Fig:eigenvalue}b), the complex conjugate pair of eigenvalues crosses the imaginary axis such that their real parts become positive, suggesting an instability of synchrony through a Hopf bifurcation. 

\section{\label{sec:discussion}Distinguished features of novel transition}

\begin{figure}[t!]
\centering
\includegraphics[width=0.7\columnwidth]{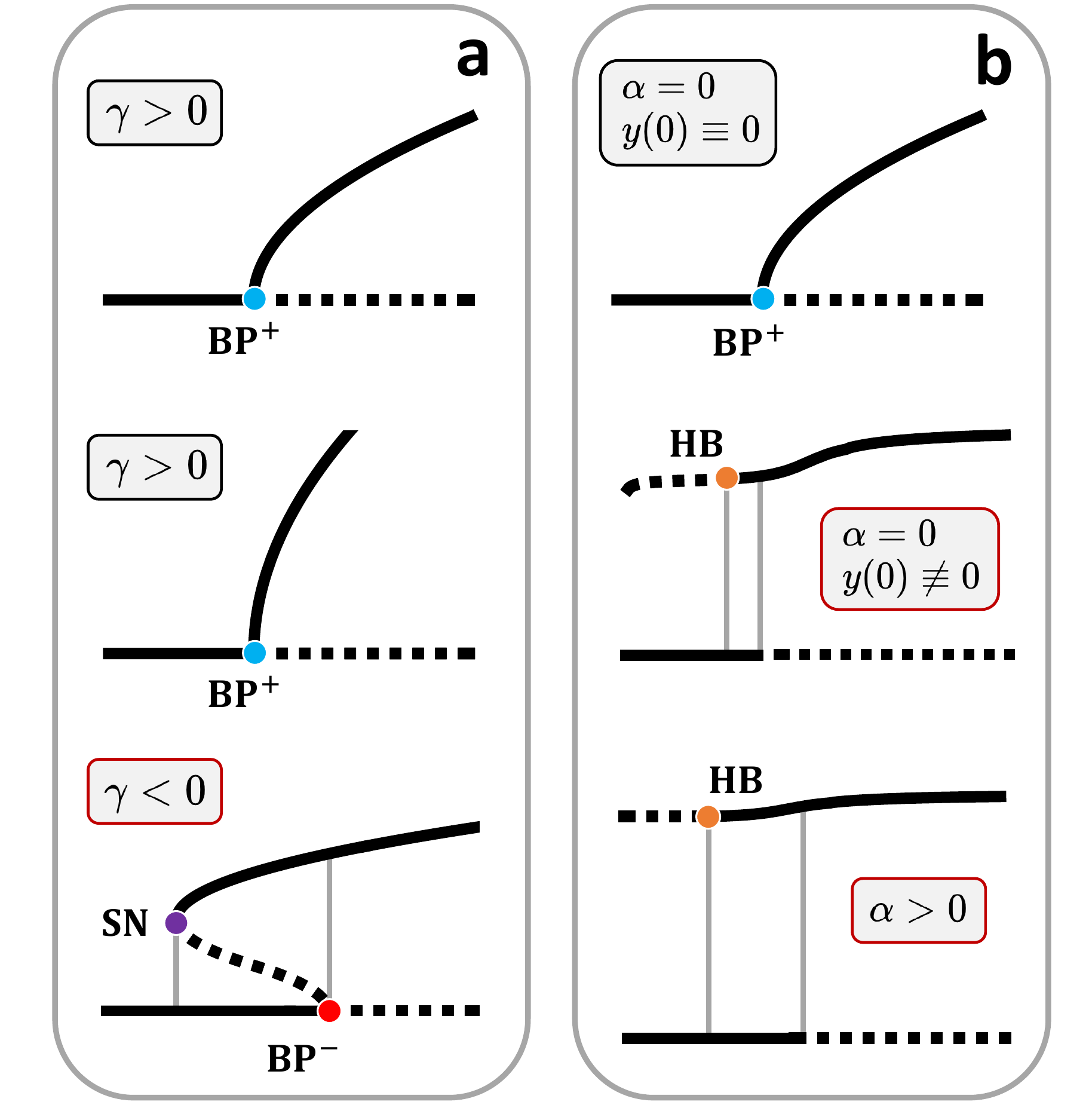}
\caption{\label{Fig:fig3} \textbf{Disconnected branches induce novel type of discontinuous transitions.} (a), (b) Schematics of the order parameter $r$ \eqref{eq:OP_kuramoto} as a function of coupling strength $|K|$. (a)  Traditionally, incoherent and synchronous states are connected, and changes from continuous to discontinuous transitions emerge through local bifurcations (e.g. branching points $\textrm{BP}^{\pm}$ such as transcritical or pitchfork) upon varying an additional parameter (here named $\gamma$), compare Ref.~\cite{kuehn2021}. As $\gamma$ varies, a branching that \textit{connects} the incoherent state with the synchronous state changes from supercritical ($\textrm{BP}^+$) to subcritical  ($\textrm{BP}^{-}$), thereby inducing hysteresis through a saddle-node (SN) bifurcation in the thermodynamic limit. (b)  Persistently discontinuous phase transitions we present emerge through two \textit{disconnected} branches (no branching point).  Top panel: Only for $\alpha = 0$  and for a measure-zero set of initial conditions ($y(0) \equiv 0$), a continuous transition occurs through a supercritical branching point ($\textrm{BP}^+$). Stable synchrony emerges via a Hopf bifurcation (HB) upon increasing the coupling strength~\cite{Lee2025}. 
For both panels, solid curves indicate a stable state while dashed curves mark unstable states.
}
\end{figure}

The discontinuous transitions to synchrony we uncovered are thus drastically different from widely known discontinuous and explosive transitions \cite{BOCCALETTI20161,d2019explosive}. Key recent work \cite{kuehn2021} found that the transition to synchrony occurring with increasing coupling strength often smoothly changes from continuous to discontinuous (or vice versa) upon varying an additional parameter, say $\gamma$. The core mechanism for such change in transition type has been identified as follows (Fig \ref{Fig:fig3}a): Varying $\gamma$ (continuously) changes the transition between subcritical and supercritical. Two examples illustrated in detail include pitchfork and transcritical bifurcations. For all discontinuous transitions arising through any such mechanism, an unstable branch remains that \textit{connects} the synchronous to the asynchronous incoherent state. Thus, established discontinuous transitions to synchrony connect the synchronous and incoherent states through an unstable branch. In traditional discontinuous transitions, stable synchronous states appear through a saddle-node bifurcation, with the associated unstable branch connecting to the incoherent state  (Fig \ref{Fig:fig3}a bottom subpanel). The reverse, a disappearence of synchrony constitutes an example of a well-known scenario of discontinuous, explosive, or catastrophic phenomena~\cite{Strogatz2000Book}. Moreover, the transition is continuous in ``half" the parameter space, say for $\gamma>0$, whereas it is discontinuous in the other half, for $\gamma<0$.

In contrast, the transition between synchrony and incoherence we uncovered (Fig.~\ref{Fig:fig3}b) exhibits intriguing novel characteristics:

\begin{enumerate}
    \item[(i)] The state of synchrony appears to be entirely \textit{disconnected} from the incoherent state.
    
    \item[(ii)] Synchrony achieves (de-) stabilization via a Hopf bifurcation rather than a saddle-node bifurcation (Fig.~\ref{Fig:eigenvalue}).

    \item[(iii)] The transitions are discontinuous \textit{almost everywhere} up to a single point in parameter space ($\alpha=0)$ and special initial conditions ($z_\mu(0) \in \mathbb{R}$ for all $\mu \in [N]$). Thus discontinuous transitions are persistent.
\end{enumerate}

\section{Summary and Outlook}

The above results thus indicate the existence of a novel transition between synchrony and incoherence that not only is generically discontinuous, but is mediated by a Hopf bifurcation that destabilizes the synchronous state. The synchronous state therefore appears to be disconnected from the incoherent state, in contrast to established synchronization transitions\cite{Strogatz2000kuramotoCrawford, Rodrigues2016kuramoto, kuehn2021}. 

Exactly determining the point at which the incoherent state becomes unstable constitutes a challenge, in particular in large systems, because the incoherent state just constitutes a complicated transient trajectory in finite systems. In the thermodynamics limit, while the quaternion order parameter we proposed has helped analyze the type of transition (continuous vs. discontinuous), as the system consists of two-variable units, we were not successful in pinning down their joint distribution functions and the associated time evolution equations. 

More generally, open challenges involve understanding potential new types of phase transitions indicating the emergence of temporal (or structural) order.   Coupled nonlinear systems with states and parameters analytically continued to become complex as in the Kuramoto model \cite{Moritz_thuemler2023, Lee2025, lee_complex} discussed here, may lead the way in exploring such challenges. Indeed, in the past, such a continuation has led to fields like PT-symmetric quantum mechanics, fractal geometry, and contributed to the foundations of  the statistical physics of phase transitions~\cite{Yang1952,Lee1952,mandelbrot2004fractals,bender1998real,benderBook2019}. For instance, recent work already provides a few hints about which new forms of collective dynamics may emerge in other complexified models such as the Winfree model~\cite{Seung_Yeal2021} or networks of neurons~\cite{rok2024,lohe2025,rokpazo2025}. A quaternion order parameter may help not only to quantify order but also to qualitatively distinguish transition types, as we have demonstrated by the example of complex locked states above. We speculate that it may, in particular, help identify further novel types of transitions that are induced by the existence of state variables in addition to phases.

\paragraph*{Acknowledgements} This work has been partially supported by the German Federal Ministry for Education and Research (BMBF) under grant number 03SF0769 (ResiNet) as well as the German National Science Foundation (Deutsche Forschungsgemeinschaft DFG) under grant number DFG/TI 629/13-1 (SynCON) and DFG/TI 629/14-1 (Koselleck) under Germany´s Excellence Strategy – EXC-2068 – 390729961, through the Cluster of Excellence \textit{Physics of Life}. 

\paragraph*{Competing interests}
The authors declare no competing interests.

\appendix

\section{\label{appendA}Quaternion order parameter and its components}

In the main text, the quaternion order parameter is defined as
\begin{flalign}
    q &:= \frac{1}{N}\sum_{\nu=1}^N (\textrm{j} ~\overline{\sin z_\nu}+ \cos z_\nu) \notag \\
    &= q_0 + q_1 \textrm{i} + q_2 \textrm{j} + q_3 \textrm{k}\in \mathbb{H}, \label{eq:QOP_suppl}
\end{flalign} where $\textrm{i}, \textrm{j}, \textrm{k}$ are basis elements of quaternions $\mathbb{H}$ and an overline indicates a complex conjugation. Using properties of complex sine and cosine functions such as $\sin(x+\textrm{i}y) = \sin x \cosh y +\textrm{i} \cos x \sinh y $ and $\cos(x+\textrm{i}y) = \cos x \cosh y -\textrm{i} \sin x \sinh y $, the real and vector part components of the quaternion order parameter \eqref{eq:QOP_suppl} are explicitly written as
\begin{flalign}
    q_0 &= \frac{1}{N}\sum_{\nu=1}^{N} \cos x_\nu \cosh y_\nu \notag \\
    q_1 &= -\frac{1}{N}\sum_{\nu=1}^{N} \sin x_\nu \sinh y_\nu  \notag \\
    q_2 &= \frac{1}{N}\sum_{\nu=1}^{N} \sin x_\nu \cosh y_\nu  \notag\\
    q_3 &=\frac{1}{N}\sum_{\nu=1}^{N} \cos x_\nu \sinh y_\nu. \label{eq:QOP_comp_supp}
\end{flalign} for complex dynamical units $z_\mu = x_\mu + \textrm{i} y_\mu$ for $\mu \in [N]$.

\section{\label{appendb}Quaternion order parameter effectively decouples equations of motion}

As the complex order parameter decouples the original Kuramoto equations, the quaternion order parameter \eqref{eq:QOP_suppl} decouples the analytically continued equations of motion such that
\begin{flalign}
    \frac{\textrm{d}}{\textrm{d}t}z_\mu &= \omega_\mu +\frac{K}{N}\sum_{\nu=1}^N \sin(z_\nu-z_\mu) \label{eq:ckm_suppl} \\  
    &= \omega_\mu - K \bigg( \frac{1}{N}\sum_{\nu=1}^N\cos z_\nu \bigg) \sin z_\mu + K \bigg( \frac{1}{N}\sum_{\nu=1}^N\sin z_\nu \bigg) \cos z_\mu  \notag \\
    &= \omega_\mu -K \sin z_\mu   \frac{1}{N}\sum_{\nu=1}^N \big( \cos x_\nu \cosh y_\nu - \textrm{i} \sin x_\nu \sinh y_\nu \big) \notag \\ 
    &~~ ~~+ K \cos z_\mu \frac{1}{N}\sum_{\nu=1}^N \big( \sin x_\nu \cosh y_\nu + \textrm{i} \cos x_\nu \sinh y_\nu \big)   \notag \\
    & = \omega_\mu - K (q_0 + \textrm{i} q_1) \sin z_\mu + K(q_2 + \textrm{i} q_3) \cos z_\mu \label{eq:decoupled_suppl}
\end{flalign} for $\mu \in [N]$. This form makes the mean-field nature of the model clear. Each oscillator, in effect, is driven by itself $z_\mu$ and the mean-field (the quaternion order parameter) $q$ of all other units.

\section{\label{appenC}Relation between classical and quaternion order parameters}

The quaternion order parameter effectively decouples complexified equations of motion as in \eqref{eq:decoupled_suppl}. Direct numerical integration of \eqref{eq:ckm_suppl} indicates that synchrony emerges as a fixed point solution to \eqref{eq:ckm_suppl}, called a complex locked state denoted by $z^*_\mu$ for $\mu \in [N]$. Thus, we assume that the quaternion order parameter \eqref{eq:QOP_suppl} also becomes stationary and denote it $q^* = q^*_0 + q^*_1 \textrm{i} + q^*_2 \textrm{j} + q^*_3 \textrm{k}$. When symmetric natural frequencies, i.e. $\omega_\mu = -\omega_{\mu+N/2}$ for $\mu \in [N/2]$, are imposed on oscillators, we assume that a complex locked state reads $z^*(\omega_\mu) = -z^*(\omega_{\mu +N/2})$ for $\mu\in[N/2]$. It then follows in \eqref{eq:QOP_comp_supp} that $q_2^* = q^*_3=0$. Finally, the complex locked state reads
\begin{flalign}
    z_\mu^* = \sin^{-1} \bigg( \frac{\omega_\mu}{K(q^*_0 + \textrm{i}q^*_1)} \bigg) =\sin^{-1} \bigg( \frac{\omega_\mu}{K q^*} \bigg)
\end{flalign} for $\mu \in [N]$. Numerical simulations also indicate the result described here.

In the thermodynamic limit, a complex locked state $z^*(\omega) = \sin^{-1} \big( \frac{\omega}{K q^*} \big)$  gives a relation between the quaternion order parameter and the classical Kuramoto order parameter. To see this, we consider
\begin{flalign}
     x^*(\omega) = \textrm{sgn}(\omega) \sin^{-1}  \sqrt{\frac{1}{2} \bigg(1 + c^2 -\sqrt{1+ c^4 - 2 c^2 \cos 2\theta  } \bigg)  } 
\end{flalign} where we define $c := \frac{\omega}{|K| \|q\|} $, and $\theta := \alpha + \arg(q)$. Then, the classical Kuramto order parameter reads
\begin{flalign}
    r &= \bigg| \int_\mathbb{R} \big( \textrm{i}\sin[ x^*(\omega) ]+ \cos [x^*(\omega)] \big) g(\omega) \textrm{d}\omega \bigg| \notag \\
    &=  \int_\mathbb{R}  \cos [x^*(\omega)]  g(\omega) \textrm{d}\omega = \int_\mathbb{R} \bigg(1 - \sin^2[x^*(\omega)]  \bigg)^{1/2}   g(\omega) \textrm{d}\omega  \notag \\
    &=\int_\mathbb{R}\Bigg( \frac{1}{2} - \frac{\omega^2}{2|K|^2 \| q\|^2}+\frac{1}{2} \sqrt{1+ \frac{\omega^4}{|K|^4 \| q\|^4} - \frac{2\omega^2 \cos 2\theta}{|K|^2 \| q\|^2}} \Bigg)^{1/2}  g(\omega) \textrm{d} \omega \label{eq:OP_QOP}
\end{flalign} in terms of $q$. In particular, as $|K| \rightarrow 0^+$, the Kuramoto order parameter in \eqref{eq:OP_QOP} becomes $r \sim \sqrt{\frac{1-\cos2\theta}{2}} + \mathcal{O}(|K|^2)$. This is consistent with the unstable complexified synchrony $r \sim \sqrt{\frac{1+\sin \alpha}{2}}$, via the relation $\arg(q) \sim \frac{\pi}{4}- \frac{\alpha}{2}$ as $|K| \rightarrow 0^+$. Also, it follows that $r \rightarrow 1$ as $|K| \rightarrow \infty$ as discussed in the main text. 

For finite-size systems, one may conclude that the classical Kuramoto order parameter is obtained via
\begin{flalign}
    r = \frac{1}{N}\sum_{\nu=1}^N \Bigg( \frac{1}{2} - \frac{\omega_\nu^2}{2|K|^2 \| q\|^2}+\frac{1}{2} \sqrt{1+ \frac{\omega_\nu^4}{|K|^4 \| q\|^4} - \frac{2\omega_\nu^2 \cos 2\theta}{|K|^2 \| q\|^2}} \Bigg)^{1/2}
\end{flalign} as long as the quaternion order parameter $q$ is measured through the self-consistency equation (in the main text).

\section{\label{appenB234}Quaternion order parameter vanishes for the incoherent states}

As in the main text, the oscillator density for the incoherent state in the thermodynamic limit is inversely proportional to the velocity $\frac{\textrm{d}}{\textrm{d}t}z$ such that
\begin{flalign}
    \rho_\textrm{inc}(x,y,\omega)
  =  \frac{C(\omega)}{
     \big|\omega -K [(q_0 +  \textrm{i} q_1) \sin(x+\textrm{i}y) +  (q_2 + \textrm{i} q_3) \cos(x+\textrm{i}y)] \big|  } \label{eq:rho_inc_suppl}
\end{flalign} where $C(\omega)$ is the normalization constant. In Eq.~\eqref{eq:rho_inc_suppl}, the oscillator density satisfies a symmetry property $\rho_\textrm{inc}(x,y,\omega) = \rho_\textrm{inc}(x+\pi,y,-\omega)$ and the natural frequency density is assumed to be even $g(\omega) = g(-\omega)$. It then follows from these two symmetries that the quaternion order parameter for the incoherent states vanishes such that
\begin{flalign}
    q_\textrm{inc} &= \int_\mathbb{R}\int_{\mathbb{S}^1} \int_\mathbb{R} \rho_\textrm{inc}(x,y,\omega) (\textrm{j}\overline{\sin(x+\textrm{i}y)}+\cos(x+\textrm{i}y)) g(\omega) \textrm{d}\omega \textrm{d} x \textrm{d} y \notag \\
    &= -\int_\mathbb{R}\int_{\mathbb{S}^1} \int_\mathbb{R} \rho_\textrm{inc}(x,y,\omega) (\textrm{j}\overline{\sin(x+\textrm{i}y)}+\cos(x+\textrm{i}y)) g(\omega) \textrm{d}\omega \textrm{d} x \textrm{d} y =0    \label{eq:q_inc_suppl}
\end{flalign} in the thermodynamic limit.

\section{\label{appenBbbbb}Detached discontinuous phase transitions to synchrony for general frequency densities}

In the main text, we exploited natural frequencies obtained from the relation $\omega_\mu = -\omega_{\mu+N/2}$ for $\mu \in [N/2]$ where the quaternion order parameter \eqref{eq:QOP_suppl} is restricted to become a complex number. Here, we provide a detour to persistently discontinuous phase transitions to synchrony, not relying on the quaternion order parameter restricted to complex numbers. The analysis below confirms the same result as in the main text.

To this end, we reformulate the complexified Kuramoto equation \eqref{eq:ckm_suppl} into a Riccati form, which then satisfies 
\begin{flalign}
    \frac{\textrm{d}}{\textrm{d} t} h_\mu = \textrm{i} \omega_\mu h_\mu +\frac{K}{2} \Gamma_1 - \frac{K}{2} \Gamma_2 h_\mu^2
\end{flalign} for $\mu \in [N]$ where we define $h_\mu := e^{\textrm{i}z_\mu}$, $\Gamma_1 := \frac{1}{N}\sum_{\nu=1}^{N} h_\nu$ and $\Gamma_2 := \frac{1}{N}\sum_{\nu=1}^{N}h_\nu^{-1}$. Following Kuramoto's self-consistency method in the thermodynamic limit, an equilibrium solution reads
\begin{flalign}
    h = \frac{\textrm{i} \omega \pm \sqrt{K^2 \Gamma_1 \Gamma_2 - \omega^2}}{K \Gamma_2} \label{eq:equili}
\end{flalign} that results from an algebraic equation $0 = \textrm{i} \omega h +\frac{K}{2} \Gamma_1 - \frac{K}{2} \Gamma_2 h^2 $. The plus sign branch in \eqref{eq:equili} leads to a self-consistency equation
\begin{flalign}
    \Gamma_1 &= \int_{-\infty}^{\infty} h(\omega) g(\omega) \textrm{d}\omega \notag \\ 
    &= \int_{-\infty}^{\infty} \frac{\textrm{i}\omega+ \sqrt{K^2 \Gamma_1 \Gamma_2 - \omega^2}}{K \Gamma_2} g(\omega) \textrm{d}\omega. 
\end{flalign} Defining a macroscopic observable $R := \sqrt{\Gamma_1 \Gamma_2} \in \mathbb{C}$, we obtain a self-consistency equation for a complexified synchrony, which reads
\begin{flalign}
    K R^2 = \int_{-\infty}^{\infty} \sqrt{K^2 R^2 -\omega^2} g(\omega) \textrm{d}\omega \label{eq:self-consistency}
\end{flalign} where $R $ characterizes well different states of order and disorder in the synchronization transition process, as the quaternion order parameter does so. 

\paragraph*{Continuous phase transition to synchrony}

When both the initial condition and the coupling constant are exactly real, i.e. $\alpha=0$ and $|h_\mu|=1$ for $\mu\in[N]$, the macroscopic order parameter becomes the magnitude of the classical Kuramoto order parameter and thus $R \in \mathbb{R}$. To yield a self-consistent real $R$, the argument of the root in the self-consistency equation \eqref{eq:self-consistency} needs to be non-negative, thereby constraining the range of frequencies $\omega$ such that
\begin{flalign}
    R &= \int_{-KR}^{KR} \sqrt{1-\frac{\omega^2}{K^2R^2}} g(\omega) \textrm{d}\omega \notag \\ 
    &=KR \int_{-1}^{1} \sqrt{1-x^2} g(K R x) \textrm{d}x \label{eq:self_real}
\end{flalign} where $x := \frac{\omega}{KR}$. Taking $R \rightarrow 0^+$, the critical coupling strength is obtained via $1 = K_c g(0) \int_{-1}^{1}\sqrt{1-x^2}\textrm{d}x = K_c g(0) \frac{\pi}{2}$, as in the main text. We thus recover the continuous phase transition to synchrony known from the original Kuramoto model.

\paragraph*{Persistence of discontinuous transitions}

Following the same logic in the main text, in general, the self-consistency equation \eqref{eq:self-consistency} for a complex locked state $KR \in \mathbb{C}\setminus \mathbb{R}$ reads as $ |R| \rightarrow 0$
\begin{flalign}
    R &=  \int_{-\infty}^{\infty} \sqrt{1 - \frac{\omega^2}{K^2R^2}} g(\omega) \textrm{d}\omega \notag \\ 
    &= g(0) \int_\mathbb{R} \bigg( 1- \frac{\omega^2}{K^2 R^2} \bigg)^{1/2} e^{\frac{f''(0)}{2}\omega^2} \textrm{d} \omega +\textrm{t.s.t} \notag \\
    &= g(0)\frac{2}{|f''(0)|} \bigg( \frac{-1}{K^2 R^2}\bigg)^{1/2} + \mathcal{O}(|R|) +\textrm{t.s.t}
    \label{eq:self_comp}
\end{flalign}  While the LHS converges to zero asymptotically as $|R| \rightarrow 0$, the leading term of the RHS diverges. Hence, the  self-consistency equation does not admit a solution with arbitrarily small $|R|$. The finding indicates that synchronous states do not connect to the incoherent state. Any phase transition to synchrony is thus necessarily discontinuous.

\section{\label{appenG}Kuramoto order parameter as $|K| \rightarrow 0^+$ and $|K| \rightarrow \infty$}

A system of complexified Kuramoto oscillators $z_\mu = x_\mu+ \textrm{i} y_\mu \in \mathbb{C}$ is governed by \eqref{eq:CKM}. The system \eqref{eq:CKM} for $\beta:= \frac{\pi}{2}-\alpha=0$ ($\alpha =\frac{\pi}{2}$; purely imaginary coupling) reads
\begin{flalign}
    \frac{\textrm{d}}{\textrm{d} t} x_\mu &= \omega_\mu - \frac{|K|}{N}\sum_{\nu=1}^{N}\cos(x_\nu-x_\mu)\sinh(y_\nu-y_\mu) \label{eq:real_pure} \\
    \frac{\textrm{d}}{\textrm{d} t} y_\mu &=  \frac{|K|}{N}\sum_{\nu=1}^{N}\sin(x_\nu-x_\mu)\cosh(y_\nu-y_\mu) \label{eq:imaginaty_pure}
\end{flalign} for $\mu \in [N]$. A complex locked state, i.e., a fixed point solution, is achieved by setting \eqref{eq:imaginaty_pure} to zero, leading to $x^{(0)}_\mu=0$ for all $\mu \in [N]$. Substituting this into \eqref{eq:real_pure}, we reach an algebraic equation
\begin{flalign}
    \frac{\omega_\mu}{|K|} = \frac{1}{N}\sum_{\nu=1}^{N}\sinh(y^{(0)}_\nu - y^{(0)}_\mu) \label{eq:fixed_pure}
\end{flalign} for $\mu \in [N]$. Assuming that $y^{(0)}_\mu = -\sinh^{-1}(b \tilde{\omega}_\mu)$ where $\tilde{\omega}_\mu := \frac{\omega_\mu}{|K|}$, the parameter $b$ is determined by
\begin{flalign}
     \frac{1}{b} = \frac{1}{N}\sum_{\nu=1}^{N}\sqrt{1+(b\tilde{\omega}_\nu)^2} \label{eq:bbb}.
\end{flalign} For details, see \cite{Lee2025}. Hence, the complex locked state for a purely imaginary coupling reads 
\begin{flalign}
z^{(0)}_\mu = x^{(0)}_\mu + \textrm{i} y^{(0)}_\mu = 0 -\textrm{i}\sinh^{-1}(b \tilde{\omega}_\mu )  \label{eq:cls_pure} 
\end{flalign} for $\mu \in [N]$.

In \cite{Lee2025}, the asymptotic series expansion up to the first order well characterizes a complex locked state for $\beta \rightarrow 0^+$:
\begin{flalign}
    z^*_\mu &= -g \beta  \tanh( y^{(0)}_\mu)  +\textrm{i}y^{(0)}_\mu + \mathcal{O}(\beta^2) 
    \notag \\
    &= g \beta \frac{ b\omega_\nu /|K|}{\sqrt{1+\big(\frac{b\omega_\mu}{|K|}\big)^2}}  -\textrm{i} \sinh^{-1}\bigg(\frac{b\omega_\mu}{|K|}\bigg) + \mathcal{O}(\beta^2)
    \label{eq:asymp_CLS_pure_supp}
\end{flalign} where 
\begin{flalign}
   g =\frac{\sum_{\nu=1}^{N}\cosh y^{(0)}_\nu}{\sum_{\nu=1}^{N}\cosh y^{(0)}_\nu + \sum_{\nu=1}^{N}\sinh y^{(0)}_\nu \tanh y^{(0)}_\nu} >0
\end{flalign} 
is a positive parameter.

It is noteworthy that, from \eqref{eq:asymp_CLS_pure_supp}, we obtain the leading-order asymptotic behavior of a complex locked state:
\begin{flalign}
    z^*_\mu \sim 
    \begin{cases}
        \frac{1}{2}\beta + \textrm{i} \frac{1}{2} \log \big( \frac{|K|^2}{4 b^2 \omega_\mu^2} \big), & \text{if} ~~ \omega_\mu>0 \\
        -\frac{1}{2}\beta - \textrm{i} \frac{1}{2} \log \big( \frac{|K|^2}{4 b^2 \omega_\mu^2} \big), & \text{if} ~~ \omega_\mu<0
    \end{cases} 
\end{flalign} for $\mu \in [N]$ as $\beta \rightarrow 0^+$ and $|K| \rightarrow 0$. Considering this asymptotic behavior, we obtain 
\begin{flalign}
    r = \bigg| \frac{1}{N} \sum_{\mu=1}^{N} e^{\textrm{i}x^*_\mu} \bigg| \sim \bigg| \frac{1}{2} \bigg( 1 + e^{\textrm{i}\beta} \bigg) \bigg| =\sqrt{\frac{1+\sin\alpha}{2}} \label{eq:smallK}
\end{flalign} as $\beta \rightarrow 0^+$ and $|K| \rightarrow 0$. Although this asymptotic behavior is derived for $\beta \rightarrow 0^+$ and $|K| \rightarrow 0$, numerical simulations suggest that \eqref{eq:smallK} remains valid for any $\alpha$. This analysis is consistent with the results in \ref{appenC} to any $\alpha \in \mathbb{P}$.

In the limit of $|K| \rightarrow \infty$, a system of complexified Kuramoto oscillators \eqref{eq:CKM} behaves like a system of identical complexified oscillators ($\omega_\mu = 0$ for all $\mu \in [N]$ in a rotating reference frame). This system has a complete synchronization solution as a fixed point: $x^*_\mu = 0$ and $y^*_\mu = 0$ for all $\mu \in [N]$. The Jacobian matrix evaluated at the complete synchronization reads
\begin{flalign}
    \bold{J} = \begin{pmatrix}
\cos\alpha & -\sin\alpha  \\
\sin\alpha & \cos\alpha 
\end{pmatrix} \otimes \tilde{\bold{J}}_1 = R(\alpha)\otimes \tilde{\bold{J}}_1
\end{flalign} where $(\tilde{\bold{J}}_1)_{\mu \nu}:= -|K|\delta_{\mu \nu} + \frac{|K|}{N}$ for $\mu,\nu \in [N]$ and $R(\alpha) \in \textrm{SO}(2)$ is a rotation matrix. Here, $\textrm{Eig}(\tilde{\bold{J}}_1) = |K| \times \{0,-1, \cdots, -1\}$ and also the eigenvalues of the rotation matrix $R(\alpha)$ are either $e^{\textrm{i}\alpha}$ or $e^{-\textrm{i}\alpha}$. Therefore, we obtain 
\begin{flalign}
   \frac{1}{|K|} \textrm{Eig}(\bold{J}) =  \{ 0,0,-e^{\textrm{i}\alpha}, -e^{-\textrm{i}\alpha}, \cdots, -e^{\textrm{i}\alpha}, -e^{-\textrm{i}\alpha} \}.
\end{flalign} Therefore, the complete synchronization $(x^*_\mu, y^*_\mu) = (0,0)$ becomes stable for $\alpha <\frac{\pi}{2}$ while unstable for $\alpha > \frac{\pi}{2}$.

\vskip 1cm 
\section*{References}
\bibliographystyle{unsrt}
\bibliography{apssamp.bib}

\begin{thebibliography}{10}
\bibitem{initial_condition}
Initial conditions $(x_\mu(0), y_\mu(0))$ for $\mu \in [N]$ are randomly, independently drawn from $[-\pi,\pi] \times [-10^{-2},10^{-2}]$.
\end{thebibliography}

\end{document}